\documentclass[aps,prl,reprint,superscriptaddress,showpacs,showkeys,longbibliography]{revtex4-1}
\usepackage{amssymb}
\usepackage{amsmath}
\usepackage{cancel}
\usepackage{fullpage}
\usepackage{color}
\usepackage{braket}
\usepackage{graphicx}
\usepackage{epstopdf}
\usepackage{verbatim}
\usepackage{siunitx}
\usepackage[hidelinks]{hyperref} 
\usepackage[toc,page]{appendix}
\usepackage[section]{placeins}

\def\siv{$\mathrm{SiV}^{-}$}
\def\nv{$\mathrm{NV}^{-}$}
\def\sivw{silicon-vacancy}

\mathchardef\mhyphen="2D

\begin{document}

\title{Narrow-linewidth homogeneous optical emitters\\ in diamond nanostructures via silicon ion implantation}

\date{\today}

\author{Ruffin E. Evans}
\thanks{These authors contributed equally.}
\affiliation{Department of Physics, Harvard University, 17 Oxford St., Cambridge, MA 02138}
\author{Alp Sipahigil}
\thanks{These authors contributed equally.}
\affiliation{Department of Physics, Harvard University, 17 Oxford St., Cambridge, MA 02138}
\author{Denis D. Sukachev}
\affiliation{Department of Physics, Harvard University, 17 Oxford St., Cambridge, MA 02138}
\affiliation{Russian Quantum Center, Business-center ``Ural'', 100A Novaya St., Skolkovo, Moscow 143025}
\author{Alexander S. Zibrov}
\affiliation{Department of Physics, Harvard University, 17 Oxford St., Cambridge, MA 02138}
\author{Mikhail D. Lukin}
\thanks{lukin@physics.harvard.edu}
\affiliation{Department of Physics, Harvard University, 17 Oxford St., Cambridge, MA 02138}



\begin{abstract}

The negatively-charged \sivw\ (\siv) center in diamond is a bright source of indistinguishable single photons and a useful resource in quantum information protocols. Until now, \siv\ centers with narrow optical linewidths and small inhomogeneous distributions of \siv\ transition frequencies 
have only been reported in samples doped with silicon during diamond growth. 
We present a technique for producing implanted \siv\ centers with nearly lifetime-limited optical linewidths and a small inhomogeneous distribution. These properties persist after nanofabrication, paving the way for incorporation of high-quality \siv\ centers into nanophotonic devices.

\end{abstract}

\maketitle

\section{Introduction}

	Coherent emitters of indistinguishable single photons are a basic ingredient in many quantum information systems\cite{OBrien2009}.
	Atom-like emitters in the solid state are a particularly appealing platform for practical quantum information because they can be scalably integrated into nanophotonic devices.
	However, no single solid-state system has yet combined high brightness of narrowband emission and a low inhomogeneous distribution of photon frequencies from separate emitters (indistinguishability) with ease of incorporation into nanophotonic structures on demand.
	For example, optically active semiconductor quantum dots can be bright and integrable into nanostructures, but have a large inhomogeneous distribution\cite{lodahl2015interfacing}.
	Nitrogen-vacancy (\nv) centers in bulk diamond\cite{doherty2013nitrogen} are bright and photostable, with a moderate inhomogeneous distribution that allows straightforward tuning of multiple \nv\ centers into resonance.
	These properties allow proof-of-principle demonstrations of quantum information protocols such as remote spin-spin entanglement generation\cite{bernien2013heralded,hensen2015loophole} and quantum teleportation\cite{Pfaff2014}.
	Further progress towards developing \nv\ based quantum devices has been hindered by low indistinguishable photon generation rates associated with the weak \nv\ zero-phonon line, a challenge that could be addressed by integrating \nv\ centers into nanophotonic structures.
	However, the optical transition frequencies of \nv\ centers are very sensitive to their local environment\cite{tamarat2006,Siyushev2013Optically}, making integration of spectrally stable emitters into nanophotonic structures a major challenge\cite{Faraon2012}.

	The negatively charged silicon-vacancy color center in diamond (\siv) has shown promise in fulfilling the key criteria of high brightness\cite{neu2011}, lifetime-limited optical linewidths\cite{Rogers2014}, and a narrow inhomogeneous distribution of optical transition frequencies\cite{sternschulte1994}. 
	The \siv\ (Fig.\ 1) has electronic states with strong dipole transitions where 70\% of the emission is in the zero-phonon line (ZPL) at \SI{737}{\nano\meter}\cite{neu2011}.
	The inversion symmetry of the \siv\ prevents first-order Stark shifts, suppressing spectral diffusion\cite{Rogers2014} and allowing indistinguishable photons to be generated from separate emitters without the need for tuning or extensive pre-selection of emitters\cite{Sipahigil2014}.
	When combined with a spin degree of freedom\cite{muller2014optical}, the \siv\ center's bright narrowband transition, narrow inhomogeneous distribution, and spectral stability make it a promising candidate for applications in quantum optics and quantum information science.

    \begin{figure}
      	\centering
		\includegraphics[scale=0.9]{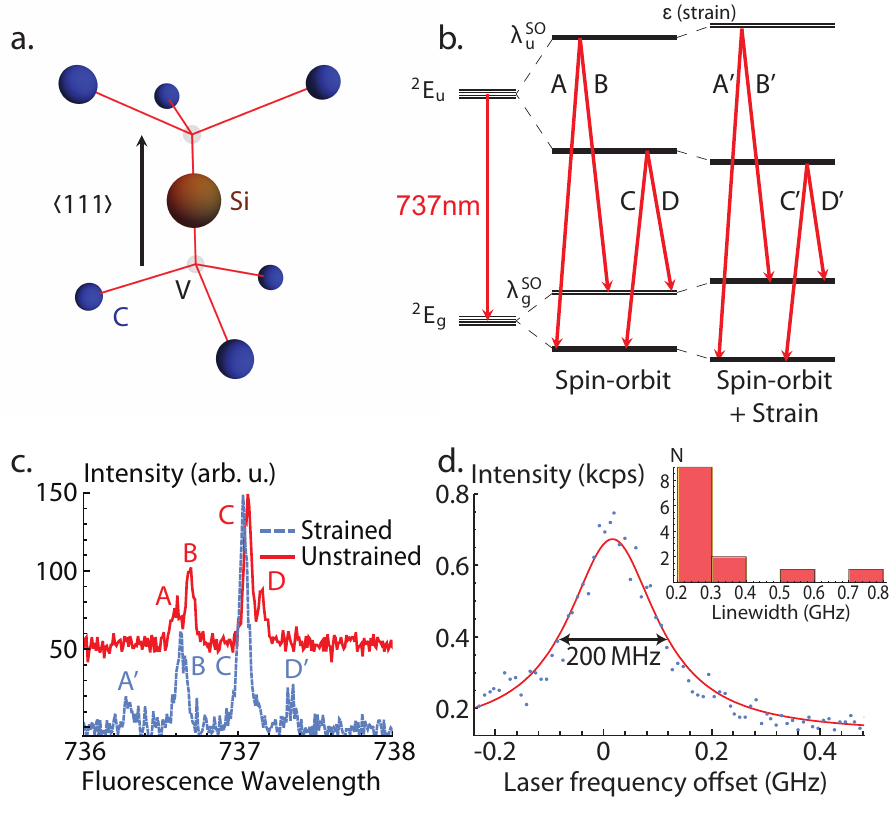}
		\caption{
			Properties of the \siv\ center.
			\textbf{a.} Atomic structure of the SiV center.
			The V-Si-V axis lies along the $\langle 111 \rangle$ lattice direction. The \siv\ has $\mathrm{D}_\mathrm{3d}$ symmetry.
			\textbf{b.} Level structure of the \siv\ center. The \siv\ is a single-hole system with double orbital and spin degeneracy. This degeneracy is partially lifted by spin-orbit coupling ($\lambda_{g}^{SO}=47\,\mathrm{GHz}$ and $\lambda_{u}^{SO}=260\,\mathrm{GHz} $\cite{Hepp2014,Rogers2014}). Lattice strain increases the splitting between these spin-orbit levels, shifting the transition frequencies.
			\textbf{c.} Fluorescence spectra of the ZPLs of single \siv\ centers in high-strain (blue, dashed) and low-strain (red) environments at 9--15\,\si{\kelvin}. Transitions B and C are less sensitive to strain compared with transitions A and D because the ground and excited states shift in the same (opposite) directions for transitions B and C (A and D)\cite{sternschulte1994}. Unstrained spectrum offset and scaled vertically for clarity.
			\textbf{d.} Linewidth (FWHM) of representative implanted \siv\ in bulk (unstructured) diamond measured by PLE spectroscopy (blue points: data; red line: Lorentzian fit). 
			Inset: histogram of emitter linewidths in bulk diamond.
			Almost all emitters have a linewidth within a factor of three of the lifetime limit (\SI{94}{\mega\hertz}).
			}
        
	    \label{fig:1}
    \end{figure}

	Silicon-vacancy centers occur only rarely in natural diamond\cite{Lo2014}, and are typically introduced during CVD growth via deliberate doping with silane\cite{Edmonds2008,DHaenens-Johansson2011} or via silicon contamination\cite{Rogers2014,neu2013,Clark1995,sternschulte1994,zhang2015hybrid}.
	While these techniques typically result in a narrow inhomogeneous distribution of \siv\ fluorescence wavelengths, these samples have a number of disadvantages. For example, the concentration of \siv\ centers can be difficult to control and localization of \siv\ centers in three dimensions is impossible.

	Ion implantation offers a promising solution to these problems.
	By controlling the energy, quantity, and isotopic purity of the source ions, the depth, concentration, and isotope of the resulting implanted ions can be controlled.
	Ion implantation is widely commercially available.
	Targeted ion implantation using a focused silicon ion beam is also possible, allowing for placement of silicon defects in all three dimensions with precision on the scale of tens of nanometers\cite{Tamura2014}.
	Despite the advantages of ion implantation, there have been conflicting results\cite{Wang2006,Tamura2014,Hepp2014} on the brightness and creation yield of \siv\ centers produced using this method and no systematic studies of the inhomogeneous distribution of \siv\ fluorescence wavelengths.
	Although there has been a single report of an implanted \siv\ with a linewidth roughly 10 times the lifetime limit\cite{pingault2014},
	to the best of our knowledge there has been up to now no consistent method for producing \siv\ centers with bright, narrow-linewidth emission using ion implantation.
	These two criteria of a low inhomogeneous distribution relative to the single-emitter linewidth and narrow single-emitter linewidth relative to the lifetime limit are essential for quantum optics applications\cite{OBrien2009,Aharonovich2011}.

	In this paper, we report the creation of \siv\ centers in diamond using ion implantation.
	Implantation is followed by high-temperature high-vacuum annealing to facilitate \siv\ formation and repair implantation-induced damage to the lattice.
	The resulting emitters have narrow optical transitions within a factor of four of the lifetime limited linewidth and a narrow inhomogeneous distribution such that half of the emitters have transitions that lie in a \SI{15}{\giga\hertz} window.
	Finally, we incorporate these \siv\ centers into nanostructures and demonstrate that their favorable optical properties are maintained even after fabrication.

\section{The \siv\ center in diamond}

	The \sivw\ color center is a point defect in diamond wherein a silicon atom occupies an interstitial position between two vacancies (Fig.\ 1a)\cite{Goss1996}.
	The \siv\ is a spin-$\frac{1}{2}$ system with ground (${}^2\mathrm{E}_g$) and excited (${}^2\mathrm{E}_u$) states localized to the diamond bandgap\cite{Rogers2014a,Goss1996,Gali2013}. Both states have double spin and orbital degeneracies partially lifted by the spin-orbit interaction (Fig.\ 1b) which splits each quartet into two degenerate doublets. The spin-orbit splittings for the ground and excited states are 0.19 and \SI{1.08}{\milli\eV} (47 and \SI{260}{\giga\hertz}), respectively (Fig.\ 1c)\cite{Rogers2014a,Hepp2014}.
	All transitions between the ground and excited states are dipole-allowed with a ZPL energy of 1.68\,eV ($\lambda=737\,\mathrm{nm}$) and an excited state lifetime of under 1.7\,ns\cite{Jahnke2015}.
	These optical transitions can have linewidths (Fig.\ 1d) comparable to the lifetime limit of \SI{94}{\mega\hertz}\cite{Rogers2014}.

	The \siv\ is sensitive to strain, which can both shift the average energy (for axial strain) and increase the splitting (for transverse strain) in the ground and excited state manifolds (Fig.\ 1b, last column)\cite{muller2014optical,Hepp2014}.
	Transitions B and C within the ZPL are relatively insensitive to transverse strain because their ground and excited states shift in the same direction: both upward for transition B and both downward for transition C (Fig.\ 1c)\cite{sternschulte1994}.
	Transition C is between the lowest energy ground and excited states which are also isolated from the phonon bath at low temperatures\cite{Jahnke2015}. This transition is therefore the most suitable for applications in quantum information science.
	
\section{Creating \siv\ centers with ion implantation}

	We create \siv\ centers using the following procedure:
	First, we begin with a polished CVD diamond (Element Six Inc., $\lbrack N\rbrack^0_S < 5\,\mathrm{ppb}$, \{100\} oriented top face).
	Previous work suggests that mechanical polishing produces a strained and damaged layer close to the surface that results in reduced mechanical stability of nanofabricated structures\cite{Burek2012}.
	We also expect that the strain introduced by mechanical polishing will lead to a larger inhomogeneous distribution of \siv\ wavelengths.
	We reduce this damage by removing 5\,$\mu$m of diamond through reactive ion etching, producing a smooth (under \SI{1}{\nano\meter} RMS roughness) surface.
	More details on this technique can be found elsewhere\cite{Burek2012,Chu2014}.
	An otherwise identical control sample was also put through the same implantation procedure but without this pre-etching step.
	We then implant ${}^{29}\mathrm{Si}^{+}$ ions (Innovion Corporation) at a dose of $10^{10}\,\mathrm{ions}/\mathrm{cm}^2$ and an energy of 150\,keV resulting in an estimated depth of $100\pm20$\,\si{\nano\meter}\cite{ziegler2010srim}.
	
	After implantation, we clean the samples using an an oxidative acid clean (boiling 1\,:\,1\,:\,1 perchloric\,:\,nitric\,:\,sulfuric acid)\cite{Hauf2011} and then perform two high-temperature high-vacuum ($\lesssim\,10^{-6}\,\mathrm{Torr}$) anneals.
	The first anneal is at \SI{800}{\celsius} for eight hours after a four-hour bake-out step at \SI{400}{\celsius}. At \SI{800}{\celsius}, vacancies are mobile\cite{Davies1992,Deak2014,Zaitsev2001} leading to the formation of \siv\ centers.
	The second anneal is the same as the first, but with an additional step at \SI{1100}{\celsius} with a two-hour dwell time. At this temperature, divacancies and other defects can also anneal out\cite{Acosta2009,Yamamoto2013}.
	For all annealing steps, we use slow temperature ramps ($\lesssim\,$\SI{35}{\celsius} per hour) to maintain low pressures in our furnace.
	This annealing procedure, inspired by previous work with \siv\cite{Clark1991,Clark1995} and \nv\cite{Acosta2009,pezzagna2010creation,Orwa2011,Chu2014} centers, both aids in the formation of \siv\ centers and also helps remove damage to the crystal lattice, reducing local strain.
	The residual graphitic carbon produced during these high-temperature anneals was removed by again performing the oxidative acid clean.
	Before each annealing step, we use X-ray photoelectron spectroscopy to verify that the surface is free of contaminants.
	
\section{Results and discussion}
	\subsection{\siv\ centers in bulk diamond}

   	We confirm that the \siv\ centers exhibit narrow-linewidth optical transitions by performing photo-luminescence excitation (PLE) spectroscopy after \SI{1100}{\celsius} annealing.
	In this experiment, we scan the frequency of a weak resonant laser (New Focus Velocity, linewidth $\Delta f\lesssim 25\,\mathrm{MHz}$ over the course of the experiment, stabilized with a High Finesse WS7 wavemeter) across transition C and monitor the fluorescence on the phonon-sideband (PSB).
	We integrate over several scans to reconstruct the time-averaged shape and position of the \siv\ ZPL (Fig.\ 1d).
	We perform these measurements in a helium flow cryostat at a sample stage temperature of \SI{3.7}{\kelvin} to avoid phonon-induced broadening of the optical transition\cite{Jahnke2015}.
	The emitters are resonantly excited below saturation to avoid power broadening.
	(See Appendix A for more experimental details.)
	We find that \siv\ centers in bulk diamond have narrow optical transitions with linewidths of $\Gamma/2\pi=320\pm180\,\mathrm{MHz}$ (mean and standard deviation for N\,=\,13 spatially resolved emitters). Almost all \siv\ centers have a linewidth within a factor of three of the lifetime limit (Fig.\ 1d, inset).
	As defined here, these linewidths include the effects of phonon broadening and all spectral diffusion that happens at any timescale during the course of the experiment (4--15 minutes).

	We characterize the inhomogeneous distribution of the implanted \siv\ fluorescence wavelengths after each annealing step via photoluminescence spectroscopy.
	To perform these measurements, we excite the \siv\ centers using off-resonant light from a \SI{700}{\nano\meter} diode laser. Off-resonant excitation at \SI{520}{\nano\meter} is also possible. Using both of these wavelengths together results in a superlinear enhancement in the observed count rate, suggesting that the \SI{520}{\nano\meter} laser may play a role in stabilizing the \siv\ charge state. 
	The resulting fluorescence is sent to a spectrometer (Horiba iHR550, \SI{0.025}{\nano\meter} resolution).
	We perform these measurements at 9--15\,\si{\kelvin}.

    \begin{figure}[t!]
    	\centering
    	\includegraphics[scale=1]{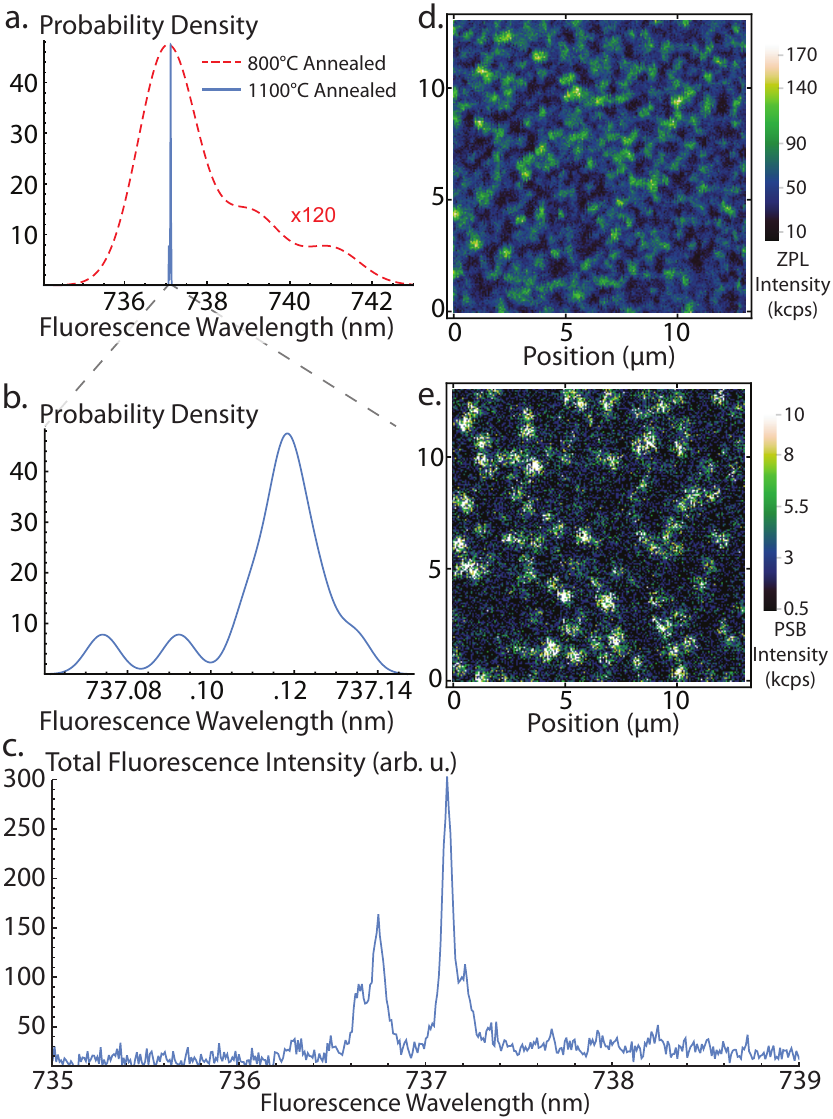}

		\caption{
			Inhomogeneous distribution of fluorescence wavelengths of implanted \siv\ transitions.
			\textbf{a.} Kernel density estimation of distribution of bulk \siv\ wavelengths after \SI{800}{\celsius} (N\,=\,19, red dashed curve) and \SI{1100}{\celsius} (N\,=\,13, blue solid curve) annealing. The distribution narrows from 3--4\,\si{\nano\meter} (\SI{800}{\celsius} anneal) to \SI{0.03}{\nano\meter} (\SI{15}{\giga\hertz}, \SI{1100}{\celsius} anneal). \textbf{b.} Zoomed-in distribution (transition C) after \SI{1100}{\celsius} annealing. Note the smaller wavelength range on the horizontal axis.
			\textbf{c.} Sum of spectra for different \siv\ centers after \SI{1100}{\celsius} annealing. The \siv\ fine structure is clearly present, demonstrating that the inhomogeneous distribution is small.
			\textbf{d}, \textbf{e.} Spatial map of collected fluorescence (thousands of counts per second) over a region of bulk diamond exciting off (d) and on (e) resonance. By comparing the densities of emitters, we estimate that $30\pm15\%$ of the emitters are nearly resonant. These measurements were taken at 9--15\,\si{\kelvin}.}
	    \label{fig:2}
    \end{figure}

	After annealing at \SI{800}{\celsius}, the observed distribution is broad, with about half of the emitter transition wavelengths lying within a 3--4\,\si{\nano\meter} range (Fig.\ 2a, red dashed curve).
	Transition C was used where unambiguous identification was possible; otherwise, the brightest transition (which should correspond to transition C\cite{sternschulte1994,Rogers2014}) was used.
	After the \SI{1100}{\celsius} anneal, the distribution becomes more than 100 times narrower, with about half of the 13 measured emitters (transition C) now lying in a \SI{0.03}{\nano\meter} (\SI{15}{\giga\hertz}) window (Fig.\ 2a and 2b, blue solid curves).
	In both cases, we focus on transition C because it is the brightest transition and relatively insensitive to strain\cite{sternschulte1994} and phononic decoherence\cite{Jahnke2015}.
	The other transitions are also much more narrowly distributed after \SI{1100}{\celsius} annealing.
	In Fig.\ 2c, we plot a composite spectrum constructed by summing over all of the normalized 13 \siv\ spectra taken after \SI{1100}{\celsius} annealing. This composite spectrum is very similar to the spectrum of a single unstrained \siv\ center (Fig. 1c) and shows the expected fine-structure splitting, demonstrating that the inhomogeneous distribution of \siv\ transition wavelengths is small compared to the fine-structure splitting.
	This result is comparable to reported inhomogeneous distributions reported for \siv\ centers created during CVD growth\cite{Rogers2014,Sipahigil2014,sternschulte1994,Hepp2014}.
	It is possible that even higher temperature annealing could further reduce this inhomogeneous distribution\cite{Clark1995,Orwa2011}.

 	To estimate the yield of conversion from implanted Si${}^+$ ions to \siv\ centers, we perform scanning confocal microscopy (Fig.\ 2d).
 	Exciting with several milliwatts of off-resonant light (\SI{700}{\nano\meter}) gives around $10^5$ counts per second (cps) into a single-mode fiber from a single \siv\ in a \SI{20}{\nano\meter} spectral range around the ZPL.
 	In the resulting microscope image, we count the number of \siv\ centers and estimate a density of \mbox{0.5--1}$/\mathrm{\mu m}^2$.
 	Based on our Si${}^+$ implantation density of $100/\mathrm{\mu m}^2$, we
	estimate our \siv\ creation yield after \SI{800}{\celsius} annealing to be 0.5--1\%.
	There was no clear difference in the yield after performing the \SI{1100}{\celsius} anneal.
	Furthermore, the yield in the sample that was not pre-etched was significantly higher (2--3\%).
	The observations that higher-temperature annealing did not increase the yield and that the sample with greater surface damage had a larger yield both support the model that \siv\ formation is limited by the presence and diffusion of nearby vacancies\cite{Clark1991,Yamamoto2013}. This yield could be increased by electron irradiating the sample to create a higher vacancy density in a controllable way\cite{Acosta2009,Clark1991,DHaenens-Johansson2011}.

	To visualize the density of nearly resonant \siv\ centers, we resonantly excited the \siv\ centers with a Rabi frequency of several GHz using an external-cavity diode laser tuned to the center of the inhomogeneous distribution.
	We scan spatially over the sample and collect fluorescence on the phonon side-band (PSB).
	The resulting image taken in the same region of the sample (Fig.\ 2e) has about a factor of three fewer emitters compared to the image taken with off-resonant excitation (N$\sim$100 vs.\ $\sim$340); roughly $30\%$ of the emitters are near-resonant (within our few GHz Rabi frequency).

\subsection{\siv\ centers in nanostructures}
	One major advantage of building quantum devices with solid-state emitters rather than trapped atoms or ions is that solid state systems are typically more easily integrated into nanofabricated electrical and optical structures\cite{Ladd2010,Vahala2003}.
	The scalability of these systems is important for practical realization of even simple quantum optical devices\cite{Li2015}.
	Unfortunately, many solid-state systems suffer serious deterioration in their properties when incorporated into nanostructures.
	For example, the large permanent electric dipole of \nv\ centers in diamond causes coupling of the \nv\ to nearby electric field noise, shifting its optical transition frequency as a function of time.
	The \siv\ is immune to this spectral diffusion to first order because of its inversion symmetry\cite{Sipahigil2014} and is therefore an ideal candidate for integration into diamond nanophotonic structures.
	Motivated by these considerations, we fabricated an array of diamond nanophotonic waveguides (Fig.\ 3a) on the pre-etched sample characterized above using previously reported methods\cite{Burek2012,Hausmann2013}.
	Each waveguide (Fig.\ 3a, inset) is 23\,$\mu$m long with approximately equilateral-triangle cross sections of side length 300--500\,\si{\nano\meter}.
	After fabrication, we again performed the same \SI{1100}{\celsius} annealing and acid cleaning procedure.
	Many \siv\ centers are visible in a fluorescence image of the final structures (Fig.\ 3b). Photon correlation measurements (Appendix B) verify our ability to create and image single \siv\ centers.

    \begin{figure}
    	\centering
    	\includegraphics[scale=0.95]{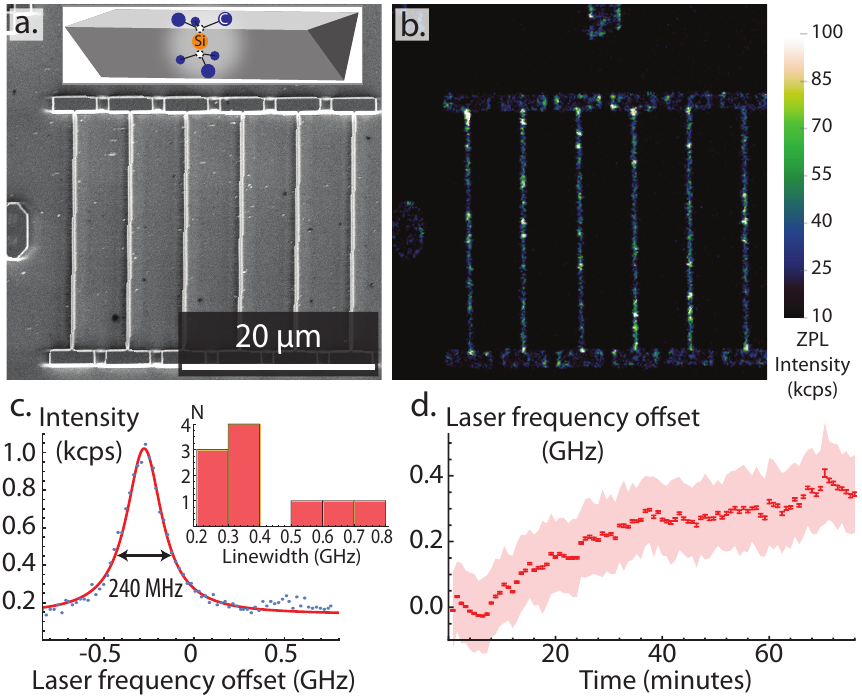}

		\caption{
			\siv\ centers in nanostructures.
			\textbf{a.} Scanning electron micrograph of six nanobeam waveguides. Inset: schematic of a triangular diamond nanobeam containing an \siv\ center.
			\textbf{b.} Spatial map of ZPL fluorescence collected by scanning confocal microscopy with off-resonant excitation. Multiple \siv\ centers are visible in each waveguide.
			\textbf{c.} Linewidth of representative implanted \siv\ inside a nano-waveguide measured by PLE spectroscopy (blue points: data; red line: Lorentzian fit).
			Inset: histogram of emitter linewidths in nanostructures. Most emitters have linewidths within a factor of four of the lifetime limit.
			\textbf{d.} Spectral diffusion of the emitter measured in part c. The total spectral diffusion is under \SI{400}{\mega\hertz} even after more than an hour of continuous measurement. This diffusion is quantified by measuring the drift of the fitted center frequency of resonance fluorescence scans as a function of time. Error bars are statistical error on the fitted center position. The lighter outline is the FWHM of the fitted Lorentzian at each time point.
		}
	    \label{fig:3}
    \end{figure}

    To characterize the optical coherence properties of \siv\ centers in nanostructures, we again perform PLE spectroscopy. \siv\ centers in nanostructures have narrow transitions with a full-width at half-maximum (FWHM) of $\Gamma_n/2\pi=410\pm 160\,\mathrm{MHz}$ (mean and standard deviation for N\,=\,10 emitters; see Fig.\ 3c inset for linewidth histogram), only a factor of $4.4$ greater than the lifetime limited linewidth $\gamma/2\pi=94\,\mathrm{MHz}$.
    The linewidths measured in nanostructures are comparable to those measured in bulk (unstructured) diamond ($\Gamma_b/2\pi=320\pm 180\,\mathrm{MHz}$).
    The ratios $\Gamma_n/\gamma$ and $\Gamma_b/\gamma$ are much lower than the values for \nv\ centers, where the current state of the art for typical implanted \nv\ centers in nanostructures\cite{Faraon2012} and in bulk\cite{Chu2014} is $\Gamma_n/\gamma\gtrsim$ 100--200 and $\Gamma_b/\gamma\gtrsim10$ ($\gamma/2\pi=13\,\mathrm{MHz}$ for \nv\ centers).
    
    It is possible for the lifetime in nanostructures to be longer than the lifetime in the bulk since the local photonic density of states is generally reduced inside such a structure\cite{Babinec2010,chu2015quantum}.
    This potential change in lifetime would change the lifetime-limited linewidth and can also provide indirect evidence of the \siv\ quantum efficiency.
	To probe this effect, we measured the lifetime of nine \siv\ centers.
	The lifetime measured in the nanobeam waveguides ($\tau=1.69\pm0.14\,\si{ns}$, N\,=\,5) was not significantly different from the lifetime measured in the bulk-like anchors ($\tau=1.75\pm0.08\,\si{ns}$, N\,=\,4). Both values are in good agreement with the literature\cite{Jahnke2015,pingault2014}. 

    By extracting the center frequency of each individual scan, we also determine the rate of fluctuation of the ZPL frequency and therefore quantify spectral diffusion (Fig.\ 3d).
    Optical transition frequencies in \siv\ centers are stable throughout the course of our experiment, with spectral diffusion on the order of the lifetime-limited linewidth even after more than an hour.
    Characterizing the inhomogeneous distribution of \siv\ centers in nanostructures is challenging because off-resonant excitation leads to strong background fluorescence, making exhaustive identification of all \siv\ centers in a given region difficult.
    Nevertheless, it is easy to find multiple \siv\ centers in nanostructures at nearly the same resonance frequency: to find the above ten emitters, we scanned the laser frequency over only a \SI{20}{\giga\hertz} range.

	The residual broadening of the optical transition can result from a combination of second-order Stark shifts and phonon-induced broadening.
	The presence of a strong static electric field would result in an induced dipole that linearly couples to charge fluctuations, accounting for the slow diffusion.
	Finally, we expect that up to \SI{50}{\mega\hertz} of additional broadening could arise from the hyperfine interaction\cite{rogers2014all} present due to our choice of ${}^{29}$Si ions.
	Determining the precise mechanisms limiting \siv\ linewidths is an important topic of future study.

	To conclude, we have presented
	optical emission from implanted \siv\ centers with a narrow inhomogeneous distribution of \siv\ optical transition wavelengths and nearly lifetime-limited optical linewidths.
	These properties persist after nanofabrication, making the \siv\ center uniquely suited for integration into quantum nanophotonic devices\cite{Aharonovich2011a,Hausmann2012}.
	Recent advances in diamond fabrication technology\cite{Burek2012,Riedrich-Moller2012,Burek2014} suggest the tantalizing possibility of scalably integrating these high-quality implanted \siv\ centers into nanowire single photon sources\cite{Babinec2010} or nanocavities\cite{Lee2012,Riedrich-Moller2014}.
	Furthermore, combining our processing procedure with targeted implantation of silicon using a focused ion beam\cite{Tamura2014} either before or after fabrication\cite{SiVCQEDToBePublished} could significantly improve photonic device yields and reproducibility by deterministically positioning individual \siv\ centers in all three dimensions.
	Our work, combined with the promise of these future advances, could make the \siv\ center a new workhorse in solid-state quantum optics.

\section*{Acknowledgments}
\begin{acknowledgments}
We thank D.\,J.\ Twitchen and M. Markham from Element Six Inc.\ for providing the electronic grade diamond samples, A.\ Sushkov and S.\ Meesala for help with annealing, and N.\,P.\ de Leon and K.\ De Greve for help with etching and sample processing. We also thank Y.\ Chu, B.\,J.\ Shields, K.\,D. Jahnke, L.\,J. Rogers, and F.\ Jelezko for discussions and valuable insight. M.\,L.\ Goldman and C.\,T.\ Nguyen helped develop some of the software used in the experiment. M.\,K.\ Bhaskar contributed to figure design. 

Financial support was provided by the NSF, 
the Center for Ultracold Atoms,
the Air Force Office of Scientific Research MURI ``Multifunctional Light-Matter Interfaces based on Neutral Atoms \& Solids'', the DARPA QuINESS program, and the ARL.
R.\,E. was supported in part by the NSF Graduate Research Fellowship Program.
This work was performed in part at the Center for Nanoscale Systems (CNS) of Harvard University which is supported under NSF award ECS-0335765.

\end{acknowledgments}

\section*{Appendix A: Experimental Setup}

	\begin{figure}[h ]
		\includegraphics[width=1.1\linewidth]{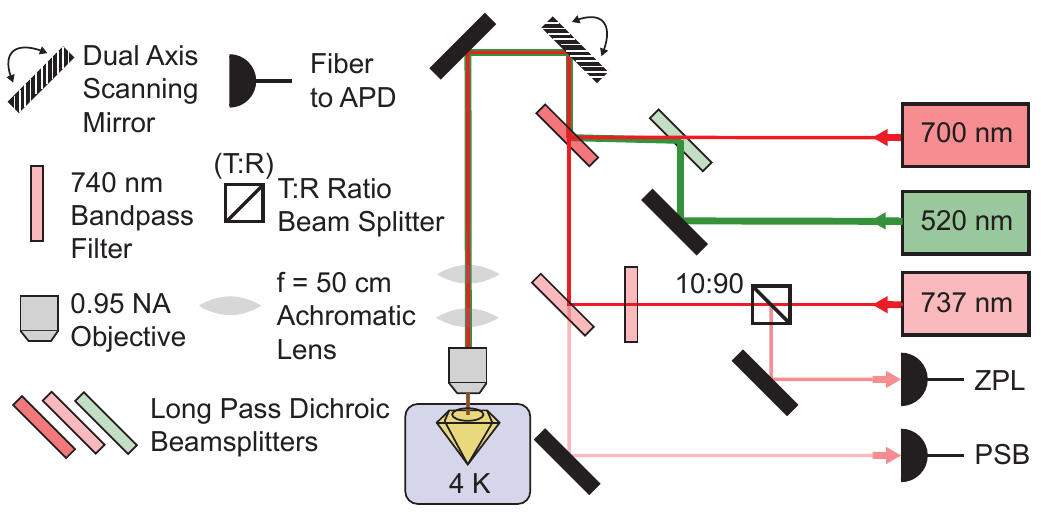}
			\caption{
			Confocal microscope design. The \SI{520}{\nano\meter} and \SI{700}{\nano\meter} lasers are used to excite the \siv\ off-resonantly. The \SI{737}{\nano\meter} external-cavity diode laser is used to excite the \siv\ resonantly. Collection can be performed either on the ZPL (if the excitation is off-resonance) or the PSB (in either excitation scheme).
			}
		\label{fig:layout}
	\end{figure}

	The experiments were carried out using home-built scanning confocal microscopes as illustrated in Fig.\ \ref{fig:layout}. The three lasers used for excitation (\SI{520}{\nm} and \SI{700}{\nm} diode lasers used for off resonant excitation, \SI{737}{\nm} external-cavity diode laser used for resonant excitation) are combined using dichroic beamsplitters. A \SI{760}{\nm} long-pass dichroic beamsplitter separates the PSB fluorescence from the rest of the optical channels. An additional bandpass filter ($740\pm13\,\si{\nm}$) is used on the ZPL channel. Single photons are detected using single photon counting modules (Picoquant $\tau$-SPAD and Excelitas SPCM-NIR). The cryogenic measurements were performed in \SI{4}{\kelvin} helium flow cryostats. We used a 0.95\,NA microscope objective (Nikon CFI LU Plan Apo Epi 100$\times$) in all experiments. During the cryogenic measurements, the objective was inside the vacuum chamber and the sample was clamped with an indium foil spacer to the cold finger of the cryostat.
	During the PLE measurements, the \SI{520}{\nm} laser is pulsed at a $\sim5\%$ duty cycle to stabilize the charge state of the \siv\ center\cite{Faraon2012,Chu2014}. The detectors are gated off during these pulses.

\section*{Appendix B: Fluorescence Autocorrelation Measurements}
		
		\begin{figure}[ht]
		\includegraphics[width=\linewidth]{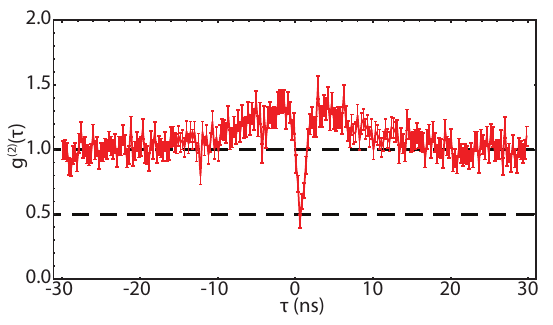}
			\caption{
			Fluorescence autocorrelation measurement of a \siv\ center inside a diamond nanobeam as described in the text. Error bars are estimated assuming the noise on the number of detected photons follows a Poisson distribution (shot noise).
			The extent of the dip at $\tau=0$ is limited by finite detector bandwidth: we measure $g^{(2)}(0)=0.45$; deconvolving the detector response yields $g^{(2)}(0)=0.15$.
			}
		\label{fig:g2}
		\end{figure}

		To verify our ability to create single \siv\ centers, we performed fluorescence autocorrelation measurements on \siv\ centers inside diamond nanobeams. We performed this measurement by exciting the \siv\ centers off resonantly as described above and splitting the emission between two detectors in a Hanbury-Brown--Twiss configuration. The relative arrival times of the photons on the two detectors were recorded using fast acquisition electronics (PicoQuant HydraHarp 400) with a resolution better than \SI{128}{\pico\second}.
		In this experiment, our total average photon count rate from this \siv\ was $9\times10^{4}$ counts per second.

		The relative photon detection times $g^{(2)}(\tau)$ (normalized by defining $g^{(2)}(\infty)=1$) from a representative \siv\ are displayed in Fig.\ \ref{fig:g2}. A value of $g^{(2)}(0)<0.5$ would confirm that we are measuring emitters producing single photons.
		Finite jitter on our detectors of around \SI{350}{\pico\second}
		causes the measured arrival times of our photons to be convolved with the detector response, hence limiting the sharpness and minimum value of our dip.
		Fitting the data (including the detector response) using previously reported methods\cite{Sipahigil2014} gives a value $g^{(2)}(0)=0.45$.
		Deconvolving the detector response gives a value $g^{(2)}(0)=0.15$, indicating that the extent of our $g^{(2)}(0)$ dip is limited primarily by detector response as expected.
		
\bibliographystyle{PRAppl_Denis}
\bibliography{refs5}

\end{document}